\documentclass[useAMS,usenatbib]{mn2e}
\usepackage{bm}
\input epsf.tex
\usepackage{graphicx}
\def\b{\bm}
\def\oline{\overline}
\title[Mean-field dynamo in partially ionized plasmas-I]
       {Mean-field dynamo in partially ionized plasmas-I}

\author[V. Krishan and R.T. Gangadhara]{V. Krishan$^{1,2}$\thanks{E-mail:
vinod@iiap.res.in} and R. T. Gangadhara$^1$ \\
$^1$Indian Institute of Astrophysics, Bangalore 560034, India\\ 
$^2$Raman Research Institute, Bangalore 560080, India}

\begin{document}

\date{Accepted--\hskip 1 cm   Received in original form --}

\pagerange{\pageref{firstpage}--\pageref{lastpage}} \pubyear{2007}

\maketitle 
 
\label{firstpage} 

\begin{abstract} 
   There are several astrophysical situations where
  one needs to study the dynamics of magnetic flux in partially
  ionized turbulent plasmas. In a partially ionized plasma the
  magnetic induction is subjected to the ambipolar diffusion and the
  Hall effect in addition to the usual resistive dissipation. In this
  paper we initiate the study of the kinematic dynamo in a partially
  ionized turbulent plasma. The Hall effect arises from the treatment
  of the electrons and the ions as two separate fluids and the
  ambipolar diffusion due to the inclusion of neutrals as the third
  fluid. It is shown that these nonideal effects modify the so called
  $\alpha$ effect and the turbulent diffusion coefficient $\beta$ in a
  rather substantial way. The Hall effect may enhance or quench the
  dynamo action altogether. The ambipolar diffusion brings in an
  $\alpha$ which depends on the mean magnetic field. The new
  correlations embodying the coupling of the charged fluids and the
  neutral fluid appear in a decisive manner. The turbulence is
  necessarily magnetohydrodynamic with new spatial and time scales.
  The nature of the new correlations is demonstrated by taking the
  Alfv\'enic turbulence as an example.

\end{abstract}

\begin{keywords}
Partially ionized plasma, Dynamo, Hall effect, Ambipolar diffusion. 
\end{keywords}

\section{Introduction}

The kinematic dynamo has revealed many an essential workings of an
astrophysical dynamo for the generation of magnetic field in objects
varying from stars to molecular clouds to accretion disks. The
kinematic dynamo (Parker 1955; Steenbeck, Krause and R\"adler 1966;
Moffatt 1970; Stix 1972) is based on the possible generation of an
electromotive force parallel to the mean magnetic field in a
reflexionally asymmetric turbulence, the so called $\alpha$ effect.
Here $\alpha$ is a measure of the net kinetic helicity. The
corresponding turbulent diffusion coefficient $\beta$ becomes a
function of the mean turbulent kinetic energy. The scale separation is
an integral part of the kinematic dynamo.  A weakly ionized plasma is
defined by the condition (Alfv\'{e}n and F\"althammer 1962) that the
electron-neutral collision frequency $\nu_{\rm en}\sim 10^{-15}n_{\rm
  n}\sqrt{8K_{\rm B}T/(\pi m_{\rm en})}$ is much larger than the
electron-ion collision frequency $\nu_{\rm ei}\sim 6\times
10^{-24}n_{\rm i}\Lambda Z^2(K_{\rm B}T)^{-3/2}$. This translates into
the ionization fraction $n_{\rm e}/n_{\rm n} < 5\times 10^{-11}T^2$
(Alfv\'{e}n and F\"althammer 1962) where $n$'s are the particle
densities and $T$ is the temperature in Kelvin. Although the ideal
magnetohydrodynamics (MHD) is often used as a starting point of an
astrophysical investigation, there are many systems with a rather low
degree of ionization dominated by the charged particle-neutral
collisions and the neutral particle dynamics. A major part of the
solar photosphere (Leake \& Arber 2006; Krishan \& Varghese 2007), the
protoplanetary disks (Krishan \& Yoshida 2006) and the molecular
clouds (Brandenburg \& Zweibel 1994) are some of the examples of
weakly ionized astrophysical plasmas. The dynamo action in such a
plasma would be affected by the multifluid interactions. The issue of
possible disconnection between the sub-surface and the surface solar
magnetic field, recently emphasized by Sch\"ussler (2005), may have
some bearing on the neglect of the neutral fluid-plasma coupling in
the flux transport on the solar photosphere. Zweibel (1988) studied
the dynamo process in a partially ionized plasma within a single fluid
description. Including only the ambipolar diffusion, she determined
the velocity, the density and the magnetic field fluctuations self
consistently in the form of magnetohydrodynamic waves and thus went
beyond the kinematic dynamo. We develop a three fluid framework for a
kinematic dynamo including the Hall effect and the ambipolar diffusion
in section two. The $\alpha$ effect of the kinematic dynamo is
formulated in section three. The new correlations arising due to the
coupling amongst different fluids are understood by taking the
Alfv\'enic turbulence as an example and we end the paper with a
section on discussion and conclusion.
\section{Three-component magnetofluid}

We begin with the three component partially ionized plasma consisting
of electrons (e), ions (i) of uniform mass density $\rho_i$ and
neutral particles (n) of uniform mass density $\rho_n.$ The equation
of motion of the electrons can be written as:
\begin{eqnarray} 
&&m_{\rm e} n_{\rm e}\left[\frac{\partial {\b V}_{\rm e}}{\partial t}+ 
({\b V}_{\rm e}\cdot\nabla){\b V}_{\rm e}\right]=
 -\nabla p_{\rm e}- \nonumber \\
&&\quad\quad e n_{\rm e}\left[{\b E}+
 \frac{\b V_{\rm e}\times \b B}{c}\right]-
 m_{\rm e}n_{\rm e}\nu_{\rm en}(\b V_{\rm e}-\b V_{\rm n})~,
\end{eqnarray} 
where the electron-ion collisions have been neglected since the
ionized component is of low density.  On neglecting the electron
inertial force, the electric field $\b E$ is found to be:
\begin{eqnarray} 
{\b E}=-\frac{\b V_{\rm e}\times \b B}{c}-\frac{\nabla
    p_{\rm e}}{en_{\rm e}}-\frac{m_{\rm e}}{e}\nu_{\rm en}(\b V_{\rm e}-\b V_{\rm n})~.
\end{eqnarray}
This gives us Ohm's law. For $\delta=(\rho_{\rm i}/\rho_{\rm n}) \ll 1$ the ion
dynamics can be ignored. The ion force balance then becomes:
\begin{eqnarray}
0=-\nabla p_{\rm i}+e n_{\rm i}\left[{\b E}+\frac
{\b V_{\rm i}\times \b B}{c}\right]-\nu_{\rm in}\rho_{\rm i}(\b V_{\rm i}-\b V_{\rm n})~,
\end{eqnarray}
where $\nu_{\rm in} $ is the ion-neutral collision frequency, and the
ion--electron collisions have been neglected for the low density
ionized component.  Substituting for $\b E$ from Eq.~(2) we find the
relative velocity between the ions and the neutrals:
\begin{equation}
\b V_{\rm n}-\b V_{\rm i}= \frac{\nabla (p_{\rm i}+p_{\rm e})}{\nu_{\rm in} 
\rho_{\rm i}}-\frac{\b J\times \b B}{ c\nu_{\rm in}\rho_{\rm i}}, 
\end{equation}
where
\begin{equation}
\b J = en_{\rm e}(\b V_{\rm i}-\b V_{\rm e})~.
\end{equation}
  The equation of motion of the neutral fluid is:
\begin{eqnarray}
  \rho_{\rm n} \left[ \frac{\partial \b V_{\rm n}}{\partial t}+({\b V_{\rm n}} \cdot\nabla )
{\b V_{\rm n}}\right]\!\!\!&=&\!\!\! -\nabla p_{\rm n}-\nu_{\rm ni}\rho_{\rm n}(\b V_{\rm n}-\b V_{\rm i})-\nonumber\\
&& \nu_{\rm ne}\rho_{\rm n}(\b V_{\rm n}-\b V_{\rm e})~, 
\end{eqnarray}
where the viscosity of the neutral fluid has been neglected.
Substituting for $\bm V_{\rm n}-\bm V_{\rm i}$ from Eq. (4), and using $\nu_{\rm in}
\rho_{\rm i}= \nu_{\rm ni} \rho_{\rm n}$ we find:
\begin{eqnarray}
\rho_{\rm n} \left[ \frac{\partial \b V_{\rm n}}{\partial t}+(\b V_{\rm n} 
\cdot\nabla)\b V_{\rm n}\right]=-\nabla p+\frac{{\b J}\times {\b B}}{c}~,
\end{eqnarray}
where $p=p_{\rm n}+p_{\rm i}+p_{\rm e}$.  
Observe that the neutral fluid is subjected to the Lorentz force as a
result of the strong ion-neutral coupling due to their collisions. 

Consider Faraday's law of induction:
\begin{equation}
\frac{\partial \b B}{\partial t}=-c\nabla\times\b E
\end{equation}
By substituting for the electric field from Eq.~(2), we get
\begin{equation}
\frac{\partial \b B}{\partial t}=\nabla\times(\b V_{\rm e}\times\b B)+
\eta \nabla^2 \b B~,
\end{equation}
where the pressure gradient terms have been dropped for the incompressible
case with constant temperature.
Here $\eta=m_{\rm e}\nu_{\rm en}c^2/(4\pi e^2n_{\rm e})$ is the electrical
resistivity predominantly due to electron-neutral collisions.
Using the construction
\begin{equation}
\b V_{\rm e}\times \b B=[\b V_{\rm n}-(\b V_{\rm n}-\b V_{\rm i})-
                        (\b V_{\rm i}-\b V_{\rm e})]\times\b B~,
\end{equation}
and substituting for the relative velocity of the ion and the neutral fluid
from Eq.~(4), Eq.~(9) becomes:
\begin{eqnarray}
{\partial \b B\over\partial t}=\nabla\times\left[\left(\b V_{\rm n} 
-\frac{\b J}{en_{\rm e}}+\frac{\b J\times\b B}{ c\nu_{\rm in}\rho_{\rm i}}
\right)\times \b{B}\right]+\eta{\nabla}^{2}\b B
\end{eqnarray}
One can easily identify the Hall term ($\b J/e n_{\rm e}$), and the
ambipolar diffusion term ($\b J\times \b B $) (Chitre \& Krishan
2001). The Hall term is much larger than the ambipolar term for large
neutral particle densities or for $\nu_{\rm in} \gg \omega_{\rm ci}$
where $\omega_{\rm ci}$ is the ion cyclotron frequency. In this system
the magnetic field is not frozen to any of the fluids. Equations (7)
and (11) along with the mass conservation
\begin{eqnarray}
\nabla\cdot\b V_{\rm n}=0
\end{eqnarray}
form the basis of our investigation. 

\section{The alpha effect in three-component magnetofluid}
The $\alpha$ effect along with its several variants is the key concept
in the generation of large scale magnetic fields from small scale
velocity and magnetic fields in the kinematic dynamo process 
(Krause \& R\"{a}dler 1980).  The magnetic induction equation (11) is 
written as:
\begin{eqnarray}
{\partial \b B\over\partial t}=\nabla\times\left[\b V_E 
\times \b{B}\right]+\eta{\nabla}^{2}\b B~,
\end{eqnarray}
where
\begin{equation}
\b V_{\rm E}= \b V_{\rm n}+\b V_{\rm H}+\b V_{\rm Am}
\end{equation}
with
\begin{eqnarray}
\b V_{\rm H}= -\frac{\b J}{en_{\rm e}}
\end{eqnarray}
as the Hall velocity and 
\begin{eqnarray}
\b V_{\rm Am}= \frac{\b J\times\b B}{ c\nu_{\rm in}\rho_{\rm i}}
\end{eqnarray}
could be called the ambipolar velocity. Following the standard
procedure (Krause \& R\"adler 1980) the velocity $\b V_{\rm E}$ and the
magnetic field $\b B$ are split into their average large scale parts
and the fluctuating small scale parts as:
\begin{eqnarray}
\b V_{\rm E} &=& \oline{\b V_{\rm E}}+\b V'_{\rm E}, \\  
       \b B &=& \oline{\b B}+\b B'
\end{eqnarray}
such that
\begin{eqnarray}
\oline{\b V'_{\rm E}}=0, \quad\quad \oline{\b B'}=0.
\end{eqnarray} 
In the kinematic dynamo the magnetic induction equation is solved for
large and small scale fields.  Substituting Eqs.~(17) and (18) into
the induction equation (11), we find, in the first order smoothing
approximation,
\begin{eqnarray}
\b V'_{\rm E}=\b V'_n-\frac{\b J'}{en_{\rm e}}+\frac{\b J'\times\oline {\b B}}
{c\nu_{\rm in}\rho_{\rm i}}
+ \frac{\oline{\b J}\times\b B'}{c\nu_{\rm in}\rho_{\rm i}}
\end{eqnarray}
and the mean flow is found to be:
\begin{eqnarray}
\oline{\b V_{\rm E}}=\oline{\b V_{\rm n}}-\frac{\oline {\b J}}{en_{\rm e}}+
             \frac{\oline{\b J}\times\oline{\b B}}{c\nu_{\rm in}\rho_{\rm i}}~.
\end{eqnarray} 

The turbulent electromotive force \mbox{\boldmath$\cal E$} is defined
as \mbox{\boldmath${\cal E}=\oline{\b V'_{\rm E}\times\b B'}$} and is a
function of the mean magnetic induction $\oline{\b B}$ and mean
quantities formed from the fluctuations. The fluctuations in
turbulence have, generally, a correlation in spatial scale $L_{cor}$ and
time scale $\tau_{\rm cor}$. In a two-scale turbulence,
$L_{\rm cor}\ll\oline L$ and $\tau_{\rm cor}\ll\oline\tau,$ where $\oline L$ and
$\oline\tau$ represent the scales of the large scale quantities. Thus
the fluctuations need to be determined in the immediate vicinity of
the point at which the large scale quantity is to be found. This
enables us to employ Taylor's expansion for $\oline{\b B}$ and express
the turbulent electromotive force as (Eq. 5.4 of Krause \&
R\"adler 1980), retaining only the first order spatial derivatives and
omitting all time derivatives,
\begin{eqnarray}
\mbox{\boldmath$\cal E$}_{\rm i} =\left(\oline{\b V'_{\rm E}\times\b B'}\right)_{\rm i} 
= a_{\rm ij}\oline{ B_{\rm j}} + b_{\rm ijk}{\partial \oline{ B_{\rm j}}\over\partial x_{\rm k}} 
\end{eqnarray}
For a zero mean flow ($\b \oline V_{\rm E}=0$) , homogeneous, isotropic,
steady and non-mirror symmetric turbulent velocity field $\b V'_{\rm
  E},$ the coefficients $ a_{\rm ij}$ and $b_{\rm ijk}$ become isotropic, and
can be expressed as:
\begin{eqnarray}
a_{\rm ij}=\alpha\delta_{\rm ij}; \\ b_{\rm ijk}=\beta\epsilon_{\rm ijk}
\end{eqnarray}
and the electromotive force can be expressed as:
\begin{eqnarray}
\mbox{\boldmath$\cal E$}= \alpha \oline{\b B} -
     \beta\, \nabla\times\oline{\b B}~,
\end{eqnarray}
The coefficients $a_{ij}$ or $\alpha$ and $b_{ijk}$ or $\beta$ become
functions of $\b\oline B$ for large $\b\oline B$ as we find here from
the contribution of the ambipolar diffusion. The quantity $\alpha$, a
pseudoscalar, turns out to be the kinetic helicity of the turbulence
and is defined as:\begin{eqnarray} \alpha &=& - \frac{\tau_{\rm
      cor}}{3}\oline{\b V'_{\rm E}\cdot
    (\nabla\times\b V'_{\rm E})}\nonumber\\
  &=& \alpha_{\rm v}+\alpha_{\rm H}+\alpha_{\rm Am}~.
\end{eqnarray}
Here, retaining terms quadratic in fluctuations,
\begin{eqnarray}
\alpha_{\rm v} = -\frac{\tau_{\rm cor}}{3}\oline{\b V'_{\rm n} \cdot\Omega'_{\rm n}}
\end{eqnarray}
is the measure of the average kinetic helicity of the neutral fluid in
the turbulence possessing correlations over time $\tau_{cor}$ and
\begin{eqnarray}
\alpha_{\rm H} = \frac{2\tau_{\rm cor}}{3en_{\rm e}}\oline{\b J'
                 \cdot\b\Omega'_{\rm n}}
\end{eqnarray}
represents the contribution of the Hall effect. The coupling of the
charged components with the neutral fluid is clearly manifest through
the possible correlation between the current density fluctuations and
the vorticity fluctuations of the neutral fluid $\b \Omega'_{\rm
  n}=\nabla\times \b V'_{\rm n} .$ The ambipolar term gives rise to
\begin{equation}
\alpha_{\rm Am} = \b \alpha_{\rm A}\cdot\oline{\b B}~,
\end{equation}
with
\begin{equation}
\b \alpha_{\rm A} = \frac{2\tau_{\rm cor}}{3c\rho_{\rm i}\nu_{\rm in}}\oline
                    {\b J'\times\b\Omega'_{\rm n}}~,
\end{equation}
as the contribution from the ambipolar diffusion with its essential
nonlinear character manifest through its dependence on the average
magnetic induction. One also observes that the Hall alpha (Eq.~28)
requires a component of the fluctuating current density along the
fluctuating vorticity of the neutral fluid whereas the ambipolar
effect (Eq.~29) thrives on the component of the fluctuating current
density perpendicular to the fluctuating vorticity. The turbulent
dissipation is given by
\begin{eqnarray}
\beta = \frac{\tau_{\rm cor}}{3}\oline{{\b V}'^2_{\rm E}}
= \beta_{\rm v}+\beta_{\rm H}+\beta_{\rm Am}
\end{eqnarray}
with
\begin{eqnarray}
\beta_{\rm v} = \frac{\tau_{\rm cor}}{3}\oline{{\b V}'^2_{\rm n}}
\end{eqnarray}
as the measure of the average turbulent kinetic energy of the neutral
fluid in the turbulence possessing correlations over time
$\tau_{\rm cor}$ and
\begin{eqnarray}
\beta_{\rm H} = \frac{2\tau_{\rm cor}}{3en_{\rm e}}\oline{\b J' \cdot\b V'_{\rm n}}
\end{eqnarray}
represents the contribution of the Hall effect. The coupling of the
charged components with the neutral fluid is clearly manifest through
the possible correlation between the current density fluctuations and
the velocity fluctuations of the neutral fluid. The ambipolar term furnishes
\begin{equation}
\beta_{\rm Am} = \b \beta_A\cdot \oline{\b B} ~,
\end{equation}
\begin{equation}
\b \beta_{\rm A} = \frac{2\tau_{\rm cor}}{3c\rho_i\nu_{\rm in}}\oline{\b 
                    J'\times\b V'_{\rm n}}
\end{equation}
with its essential nonlinear character manifest through its dependence
on the average magnetic induction. One also observes that the Hall
$\beta_{\rm H}$ requires a component of the current density
fluctuations along the velocity fluctuations of the neutral fluid
whereas the ambipolar effect thrives on the component of the current
density fluctuations perpendicular to the velocity fluctuations. We
have used rigid or perfectly conducting boundary conditions (all
surface contributions vanish) while determining the averages. Here we
consider what is known as the $\alpha^2$ dynamo and take the mean flow
$\oline{\b V_{\rm E}}=0$. This actually determines the relative mean
flow amongst the three fluids. The dynamo equation reduces to
\begin{equation}
{\partial \b B\over\partial t}=\nabla\times\left[\alpha{\b B}
-\beta\nabla\times {\b B}\right]+ \eta{\nabla}^{2}{\b B}~.
\end{equation}
From now onwards we omit the bar on the large scale magnetic field.
Assuming one dimensional space dependence, we assume the magnetic
induction $\b B=(0,~B, ~\partial A/\partial x)$ in Cartesian
coordinates $(x,~y,~z)$. In the corresponding spherical configuration,
one identifies the Cartesian coordinates $(x,~y,~z)$ with the polar
coordinates $(\theta,~ \phi, ~r)$. Thus $B$ and ${\partial A/\partial
  x}$ represent the toroidal and the poloidal parts of the field,
respectively (Stix 1972). The boundary conditions then turn out to be
the vanishing of $B$ and $A$ at the endpoints of a finite
$x$-interval, say $x=0$ and $x=\pi R$ corresponding to the poles of
the sphere.  It is convenient to put the induction equation in a
dimensionless form using a normalizing magnetic field $B_0,$ a spatial
scale $R,$ a time scale $R^2/\eta_1$ and writing $A=A' R.$ It begets:
\begin{eqnarray} 
{\partial B\over\partial t}\!\!\!\!& =&\!\!\! - R_{\alpha} {\partial^2
    A\prime\over\partial x^2}+ {\partial^2 B\over\partial x^2} -
  R_{\alpha\rm A} {\partial\over\partial x}\left[\!\left (B+a 
{\partial A'\over \partial x}\right){\partial A'\over\partial x} \right]+
       \nonumber \\
&& R_{\beta\rm A} {\partial\over \partial x}\left[\left 
(B+b {\partial A'\over\partial x}\right){\partial
      B\over\partial x}\right]
\end{eqnarray}
and
\begin{eqnarray} 
{\partial A'\over\partial t}&=& R_{\alpha}B +
  {\partial^2 A'\over\partial x^2} + R_{\alpha\rm A} \left(B+a {\partial
    A'\over\partial x}\right)B + \nonumber \\
&&R_{\beta\rm A}\left(B+b {\partial
    A'\over\partial x}\right){\partial^2 A'\over\partial x^2}~,
\end{eqnarray}
where
\begin{equation}
  R_{\alpha}   = \frac{\alpha_1 R}{\eta_1} , 
\end{equation}
\begin{equation}
  R_{\alpha\rm A} = \frac{\alpha_{\rm Ay} R B_{\rm 0}}{\eta_1}, 
\end{equation}
\begin{equation}
  R_{\beta\rm A}  = \frac{\beta_{\rm Ay} B_{\rm 0}}{\eta_1},
\end{equation}
\begin{equation}
  a = \frac{\alpha_{\rm Az}}{\alpha_{\rm Ay}},
\end{equation}
\begin{equation}
  b = \frac{\beta_{\rm Az}}{\beta_{\rm Ay}},
\end{equation}
\begin{equation}
  \eta_1 = \eta + \beta_{\rm v} + \beta_{\rm H}, 
\end{equation}
\begin{equation}
  \alpha_1 = \alpha_{\rm v} + \alpha_{\rm H}~.
\end{equation}
Here one observes that since the Hall effect contributes linearly, it
can be combined with the standard $\alpha_{\rm v}$ effect. The ambipolar
effect is nonlinear and appears separately in the induction equation.
The Hall effect can completely quench or enhance the standard
$\alpha_{\rm v}$ contribution to the dynamo for $\b V'_{\rm n} =\pm 
\b J'/(en_{\rm e})$.
In the absence of the ambipolar effect one recovers the standard
$\alpha^2$ effect with an exponential growth rate of the magnetic
induction.

It is instructive to examine the new correlations for the case of say
Alfv\'enic turbulence. Now in the weakly ionized case the relation
between the velocity and the magnetic field fluctuations for the
Alfv\'en waves is given by $\b V'_{\rm n} = \pm\delta {\b
  B'}/\sqrt{4\pi\rho_{\rm i}}$ (Krishan \& Varghese 2007) with $\delta
= \rho_{\rm i}/\rho_{\rm n}.$ Substituting these results in the
expression for $\alpha_{\rm H}$ we find:
\begin{eqnarray}
\alpha_{\rm H} =\pm \frac{2\lambda_{\rm H}}{\delta\lambda_{\rm n}}\alpha_{\rm v}
\end{eqnarray}
where $\lambda_{\rm H} = {c}/{\omega_{\rm pi}}$ is the ion inertial
scale and $\omega_{\rm pi}=\sqrt{4\pi n_{\rm i} e^2/m_{\rm i}}$ is the
ion plasma frequency and
\begin{eqnarray}
\lambda_{\rm n} = \frac{\oline{\b\Omega'_{\rm n}\cdot\b V'_{\rm n}}}
                  {\oline{\Omega'^2_{\rm n}}}
\end{eqnarray}
is the ratio of the average kinetic helicity and the average enstrophy
of the neutral fluid turbulence.  It is interesting to note that the
ambipolar $\alpha$ effect vanishes for the Alfv\'enic turbulence.  In
the next section we discuss the results of the numerical solutions of
Eqs.~(37) and (38).

\section{Discussion and conclusion}
We solve the field Eqs.~(37) and (38) demonstrating the linear and the
nonlinear $\alpha$ effect for the initial conditions given by Stix
(1972): $A'(x,~0)= 0,$ $B(x,~0)=\sin x.$ In the absence of the
ambipolar contribution ($R_{\alpha\rm A}=0, R_{\beta\rm A}=0$). The
Eqs. (37) and (38) become linear with a solution of the form
$\exp[i(kx+\gamma t)],$ where $\gamma = -k^2\pm k R_\alpha$ and $k$ is
the dimensionless wavenumber.  Then the field grows for $k<R_\alpha$
i.e., for large values of the effective $\alpha$ at large spatial
scale.

With the inclusion of the ambipolar contributions equations become nonlinear.
In Fig.~1, we present a case with the Hall and the ambipolar
contributions. The comparable values of the coefficients of the linear
and the nonlinear $\alpha$ terms with $R_{\alpha}= 1.6,$ $R_{\alpha
  \rm A}= 1.7,$ $R_{\beta\rm A}= 0.1, $ $a=1$ and $b= 0.7$ lead to a near
constant toroidal field near $ x= 2,$  fast decaying solution at
$x=0.4$ and growing solution at $x= 0.75$ beyond $t\sim 0.5$. The
poloidal field, however, grows at all values of $x$. The panels c and
d demonstrate the expected formation of spatially sharp magnetic
structures due to the nonlinearity of the ambipolar diffusion 
(Brandenburg \& Zweibel 1994). The
toroidal field in addition undergoes a reversal at $x\sim 0.5$.
Figure~2 shows the dominant effect of the ambipolar term with
$R_{\alpha}= 0.2,$ $R_{\alpha\rm A}= 3.5,$ $R_{\beta\rm A}= 1.5,$ $a=1$ and
$b= 1$. Both the components of the magnetic field, after an initial
near steady state, grow rather fast and again the formation of small
spatial scale structures is evident. Thus the inclusion of the Hall
and the ambipolar effects opens up a range of possible profiles of the
magnetic field.

In this first attempt, a framework and some instructive examples of
the dynamo solutions in a 3-component magnetofluid have been given. In
subsequent work we plan to investigate the role of the Hall and the
ambipolar terms in some realistic situations such as the Solar
surface, molecular clouds and the accretion disks. In order to deal
with these systems, the differential rotation in the objects must be
included. In the three component system, one would have to specify the
rotation profile of all the components since the system can afford to
carry a net current density. The inclusion of the ion-neutral
collisions introduces an additional time scale with which the
turbulence correlation time needs to be contrasted. The inclusion of
the Hall effect brings in the physics at the ion-inertial spatial
scale and ion gyration time scale.  The possibilities are many and
varied and should be explored in a system specific manner.

\section*{Acknowledgments}
We thank the referee for useful suggestions, and also
acknowledge the discussions with Dr. P. Subramanian during the 
initial stages of this work.

\begin{figure*}
\centerline{\epsfysize 11truecm {\epsffile[91 312 794 722]{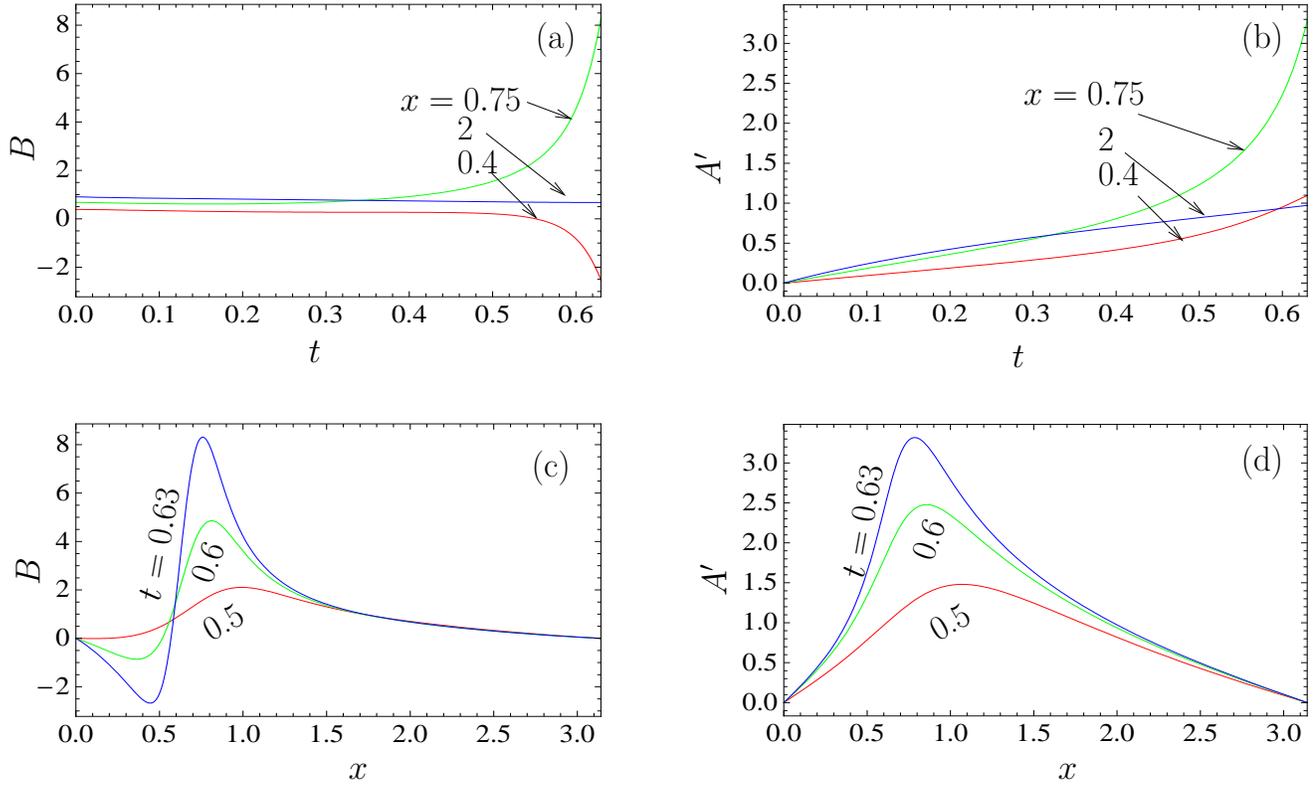}}}
\caption{Fields $B$ and $A'$ as functions of $t$ and $x.$ Panels (a)
  and (b) show their time variations at fixed $x$ values, while the
  panels (c) and (d) show the spatial variations at some fixed $t$ values.
  Chosen $R_\alpha=1.6,$ $R_{\alpha\rm A} = 1.7,$ $R_{\beta\rm
    A}=0.1,$ $a=1$ and $b=0.7.$ }
\end{figure*}
\begin{figure*}
\centerline{\epsfysize 11truecm {\epsffile[91 267 794 722]{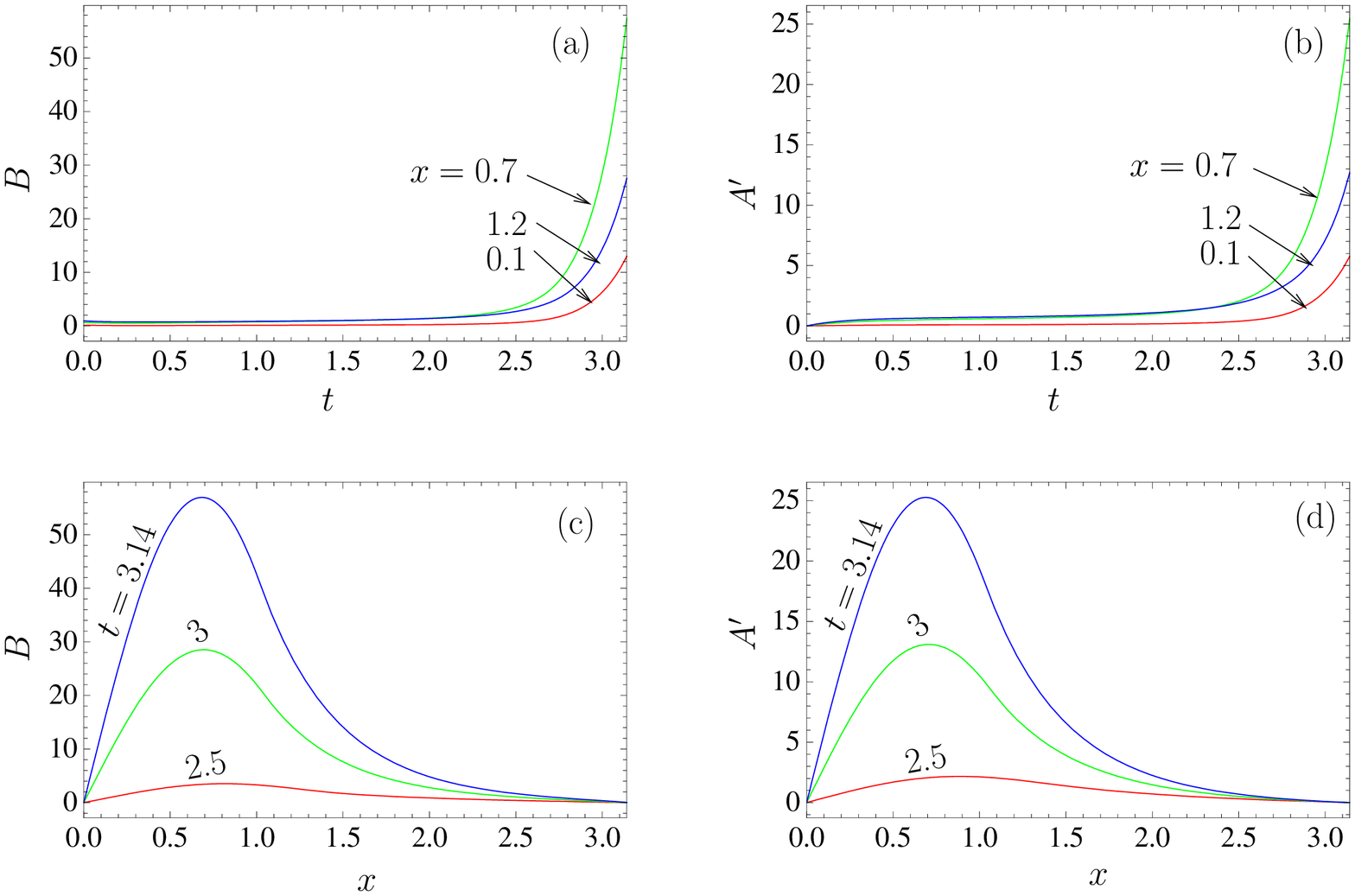}}}
\caption{Fields $B$ and $A'$ as functions of $t$ and $x.$ Panels (a)
  and (b) show their time variations at fixed $x$ values, while the
  panels (c) and (d) show the spatial variations at some fixed $t$ values.
   Chosen $R_\alpha=0.2,$ $R_{\alpha\rm A} = 3.5,$ $R_{\beta\rm
    A}=1.5,$ $a=1$ and $b=1.$ } 
\end{figure*} 
\end{document}